\documentclass[twocolumn,english,aps,prb,superscriptaddress,bibnotes,amsmath,amssymb,floatfix]{revtex4-1}
\usepackage[colorlinks=true,citecolor=blue,linkcolor=magenta]{hyperref}

\usepackage[markup=nocolor, authormarkupposition=left]{changes} 
\usepackage{soul}
\usepackage[utf8]{inputenc}
\usepackage[english]{babel}
\usepackage{amsmath,amsfonts,amssymb}
\usepackage[T1]{fontenc}
\usepackage{url}

\usepackage{amsmath}
\usepackage{amsfonts}
\usepackage{amssymb}
\usepackage{dcolumn}
\usepackage{tabularx}
\usepackage{booktabs}

\usepackage[font={footnotesize},labelfont=bf,justification=centerlast]{caption}

\usepackage{epstopdf}
\usepackage{graphicx}

\newcommand{\SiN}{Si$_3$N$_4$ }

\usepackage{changes}

\begin{document}
\title{Photonic chip-based continuous-travelling-wave parametric amplifier}

\author{Johann Riemensberger}
\email[]{johann.riemensberger@epfl.ch}
\affiliation{Institute of Physics, Swiss Federal Institute of Technology Lausanne (EPFL), CH-1015 Lausanne, Switzerland}

\author{Junqiu Liu}
\affiliation{Institute of Physics, Swiss Federal Institute of Technology Lausanne (EPFL), CH-1015 Lausanne, Switzerland}

\author{Nikolai Kuznetsov}
\affiliation{Institute of Physics, Swiss Federal Institute of Technology Lausanne (EPFL),
CH-1015 Lausanne, Switzerland}
\affiliation{Russian Quantum Center (RQC), 143026 Skolkovo, Russia}

\author{Jijun He}
\affiliation{Institute of Physics, Swiss Federal Institute of Technology Lausanne (EPFL), CH-1015 Lausanne, Switzerland}

\author{Rui Ning Wang}
\affiliation{Institute of Physics, Swiss Federal Institute of Technology Lausanne (EPFL), CH-1015 Lausanne, Switzerland}

\author{Tobias J. Kippenberg}
\email[]{tobias.kippenberg@epfl.ch}
\affiliation{Institute of Physics, Swiss Federal Institute of Technology Lausanne (EPFL), CH-1015 Lausanne, Switzerland}

\maketitle

%%%%%%%%%%%%%%%%%%%%%%%%%%%%%%%%%%%%%%%%%%%%%%%%%%%%%%%%%%%%%%%%%%%%%%
%%%%%%%%%%%%%%%%%%%%%%%%%% A B S T R A C T %%%%%%%%%%%%%%%%%%%%%%%%%%%
%%%%%%%%%%%%%%%%%%%%%%%%%%%%%%%%%%%%%%%%%%%%%%%%%%%%%%%%%%%%%%%%%%%%%%

\noindent\textbf{
The ability to amplify optical signals is of pivotal importance across science and technology. 
The development of erbium-doped-fiber-based optical amplifiers\cite{Mears:87,Desurvire:87} has revolutionized optical communications, which are today ubiquitously used in virtually all sensing and communication applications of coherent laser sources. 
In the telecommunication bands, amplifiers typically utilize rare-earth-doped fibers, or gain media based on III-V~semiconductors for integrated waveguides. 
Another way to amplify optical signals is to utilize the Kerr nonlinearity of optical fibers or waveguides via parametric processes\cite{Stolen:82, Hansryd:02}.
Such parametric amplifiers of travelling continuous wave have been originally developed in the microwave domain\cite{Yamamoto:08, Clerk:10}, where they enable quantum-limited signal amplification\cite{Clerk:10, Macklin:15} for superconducting qubit readout with high peak gain, broadband gain spectrum tailored via dispersion control, and ability to enable phase-sensitive (i.e. noiseless) amplification. 
%While parametric traveling wave amplifiers are routinely used today in the microwave domain, enabling quantum limited readout of superconducting qubits, 
Despite these advantages, optical parametric amplifiers have proven impractical in silica fibers due to the low Kerr nonlinearity. 
Recent advances in photonic integrated circuits\cite{Moss:13, Gaeta:19} have revived interest in parametric amplifiers due to the significantly increased Kerr nonlinearity in various photonic integrated platforms\cite{Foster:06, Lamont:08, Kuyken:11, Morichetti:11, Wang:15_SiH, Pu:18, Ooi:17, liu2010mid, gajda2012highly}. 
Yet, despite these major progresses\cite{Foster:06, Lamont:08, Kuyken:11, Morichetti:11, Wang:15_SiH, Pu:18, Ooi:17, liu2010mid, gajda2012highly}, all examples of net gain have necessitated pulsed lasers, limiting their practical use. 
A photonic chip-based, continuous-wave-pumped parametric amplifier, capable of amplifying arbitrary input signals, has to date remained out of reach.
Here we demonstrate a chip-based travelling-wave optical parametric amplifier with net signal gain in the continuous-wave regime. 
Using ultralow-loss, dispersion-engineered, meter-long, \SiN photonic integrated circuits\cite{Liu:21} that are tightly coiled on a chip of 5$\times$5 mm$^2$ size , we achieve a continuous parametric gain of 12~dB that exceeds both the on-chip optical propagation loss and fiber-chip-fiber coupling losses in the telecommunication C-band. 
Our work demonstrates the potential of photonic chip-based parametric amplifiers that have lithographically controlled dispersion and quantum-limited performance, and can operate in the wavelength ranges from visible to mid-infrared and outside of conventional rare-earth amplification bands. 
With future reduction of optical losses, these parametric amplifiers can be integrated or packaged with semiconductor lasers.
}

The ability to amplify optical signals is of paramount importance across science and technology. 
While optical fibers have been an instrumental development for optical communications, the choice of 1550 nm wavelength (the C- and L-band) followed the development of erbium-doped fiber amplifiers (EDFA)\cite{Mears:87,Desurvire:87}. 
The invention of EDFA has revolutionized optical communications by replacing electrical signal regeneration and enabling optical signals to propagate over more than 12000 km\cite{Temprana:15}. 
This lead to a major increase in communication bandwidth at low cost, critical to the development of the world wide web as we know it today. 
Optical amplification can also be achieved using the third-order $\chi^{(3)}$ (i.e. Kerr) nonlinearity of fibers and waveguides\cite{Stolen:82, Hansryd:02} via the parametric process. 
Such parametric amplifiers have been originally developed in the microwave domain\cite{Yamamoto:08,Clerk:10, Macklin:15}, where the term ``parametric'' designates the variation of system parameters, such as the capacitance of a transmission line or the refractive indices of optical materials. 

Parametric amplifiers have a number of unique properties that distinguish them from amplifiers based on optical transitions. 
Parametric amplifiers can achieve gain in virtually any wavelength window.
The gain can be broadband and is determined uniquely by the waveguide dispersion, leading to gain by waveguide ``design''. 
This makes parametric amplifiers attractive candidates to achieve gain in wavelength ranges that are not covered by conventional gain media. 
Parametric amplifiers operate close to the fundamental quantum noise limit of 3~dB for a single tone\cite{Hansryd:02}, and can also be operated in the phase-sensitive configuration, allowing noiseless amplification\cite{Tong:11}.  
In addition, they can have variable gain and are inherently non-reciprocal, i.e. the amplification is uni-directional. 
These properties have made parametric amplifiers pivotal for signal regeneration and wavelength conversion, and the most promising candidates to extend optical communication systems to new wavelength ranges\cite{Marhic:15}. 
Yet, despite these promises and pioneering achievements of net continuous and broadband gain\cite{Hansryd:01}, the use of parametric amplifiers has been severely limited today by the low Kerr effective nonlinearity and fabrication tolerances of optical fibers. 
In contrast, the large nonlinearity of Josephson junctions have led to the development of compact travelling-wave parametric amplifiers (TWPA) in the microwave domain\cite{Yamamoto:08,Clerk:10, Macklin:15} that are quantum-limited, exhibit broadband gain, and enable single-shot superconducting qubit readout and measurements of quantum jumps, relevant to quantum information processing\cite{Devoret:13}. 

%%%%%%%%%%%%%%%%%%%%%%%%%%%%%%%%%%%%%%%%%%%%%%%%%%%%%%%%%%%%%%%%%%%%%%
%%%%%%%%%%%%%%%%%%%%%%%%%% F I G U R E 1  %%%%%%%%%%%%%%%%%%%%%%%%%%%%
%%%%%%%%%%%%%%%%%%%%%%%%%%%%%%%%%%%%%%%%%%%%%%%%%%%%%%%%%%%%%%%%%%%%%%
\begin{figure*}[t!]
\includegraphics[width=\linewidth]{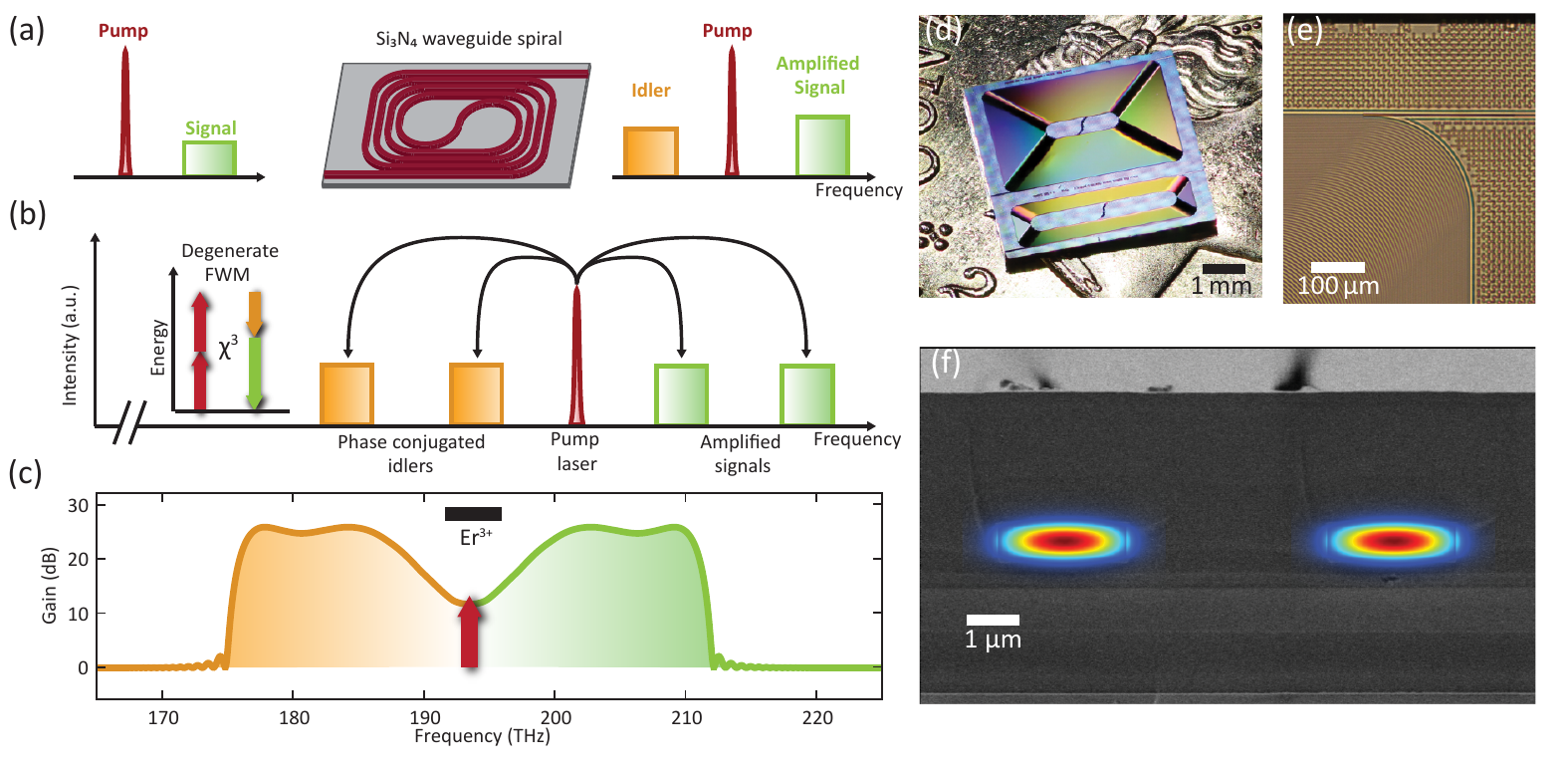}
\caption{
\textbf{Principle of a photonic chip-based, continuous-travelling-wave optical parametric amplifier (TWOPA)}.
(a)~Pump (red) and signal (green) lasers are injected into the optical waveguide spiral. 
At the output, the signal is amplified and a phase-conjugated idler is generated.
(b)~Degenerate four-wave mixing transfers optical power from the pump laser to the signal (gain) and a phase-conjugated idler (frequency conversion) that is generated at a symmetric frequency offset from the pump. 
(c)~Simulation of the maximum bandwidth for signal amplification and idler generation in a 2-meter-long, dispersion-optimized, \SiN photonic integrated waveguide. 
The achievable amplification bandwidth in \SiN waveguide spirals significantly exceeds the gain bandwidth of erbium (Er$^{3+}$) doped fiber amplifiers (black bar).  
(d)~Photograph of a \SiN photonic chip containing two waveguide spirals of lengths more than a meter. 
(e)~Optical microscope image showing a waveguide spiral corner and waveguide taper.
(f)~Scanning electron microscope image of the chip cross-section showing two parallel \SiN waveguides in the spiral.  
Simulated TE$_{00}$ mode profiles are also shown.  
}
\label{Fig:1}
\end{figure*}

\begin{figure*}[t!]
\includegraphics[width=\linewidth]{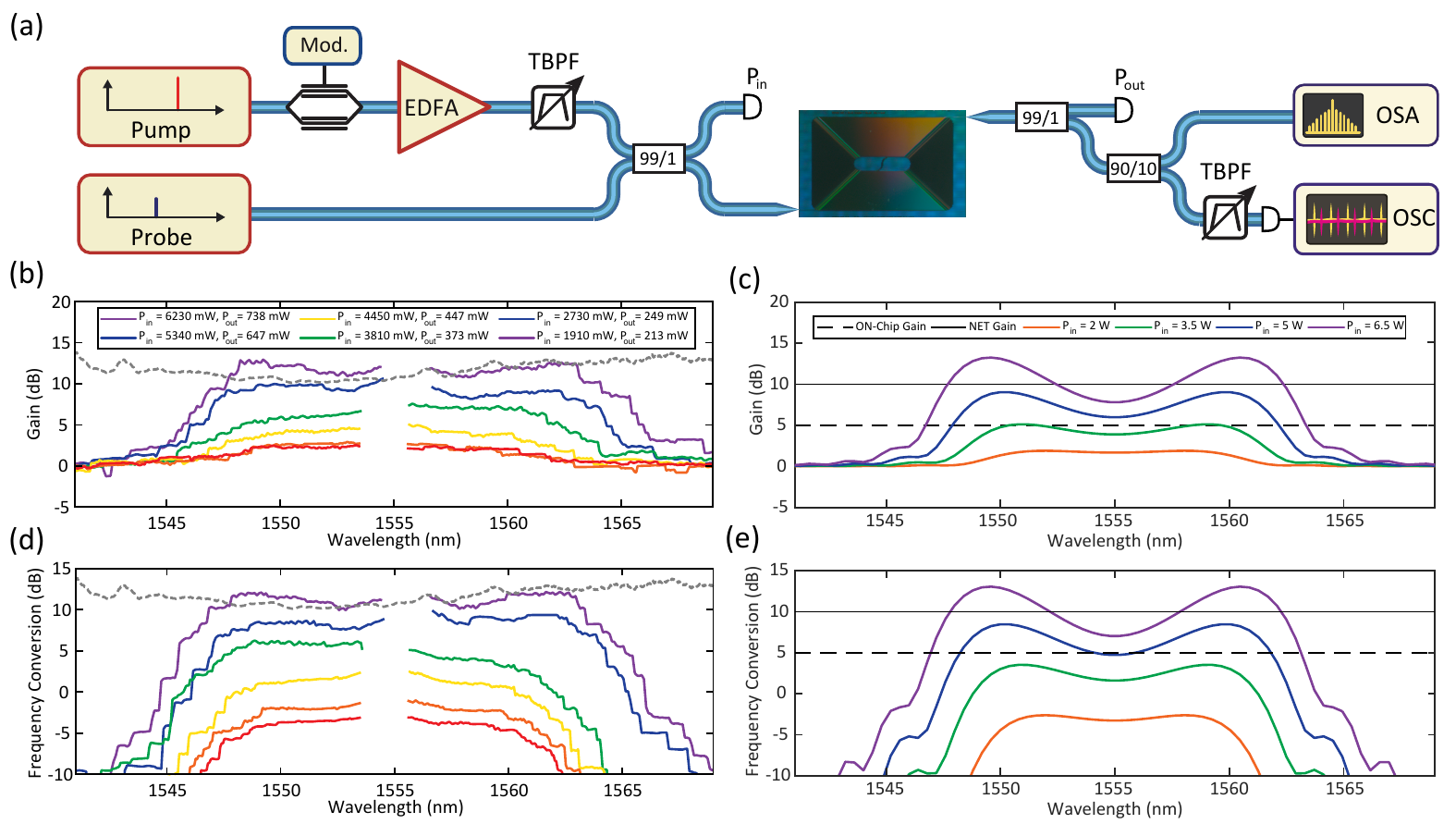}
\caption{
\textbf{Photonic chip-based, continuous-travelling-wave optical parametric amplification and frequency conversion}.
(a)~Schematic of experimental setup. 
The pump laser can be modulated (Mod.) in phase and amplitude before amplification in a high-power EDFA. 
%an optical circulator (CIRC) and beam dump (BD) suppress the back-reflection from the chip into the EDFA and amplified spontaneous emission noise is suppressed using a tunable bandpass filter (TBPF). 
An optical circulator and a beam dump are used to suppress the back-reflection from the chip into the EDFA, which are not shown here. 
Amplified spontaneous emission noise is suppressed using a tunable bandpass filter (TBPF). 
The pump and signal lasers are combined using a 20~dB directional coupler (99/1) before coupling into the photonic chip using lensed fibers and inverse tapers. 
The output light is sent to an optical spectrum analyzer (OSA). 
The amplified and modulated signal light is filtered, detected with a photodiode, and analyzed using a digital oscilloscope (OSC). 
(b)~Wavelength and power dependent gain of the system. 
Measured input ($P_\textrm{in}$) and output powers ($P_\textrm{out}$) using two power meters are marked in the legend. 
%The propagation and total loss (including fiber-to-chip coupling) of the waveguide are plotted as dashed grey and black lines, respectively.
Dashed grey line is the total loss including the fiber-chip coupling losses (two facets) and optical propagation loss in the \SiN waveguide spiral. 
(c)~Simulated gain spectra for the 2-meter-long waveguide spiral. 
Black dotted line indicates threshold for on-chip gain. 
Black solid line indicates threshold for off-chip gain.
%(d,e)~Same as (b,c) but indicating the frequency conversion efficiency from input signal to idler.
(d,e)~Same as (b,c) but indicating the frequency conversion efficiency from the pump to the idler.
}
\label{Fig:2}
\end{figure*}
%%%%%%%%%%%%%%%%%%%%%%%%%%%%%%%%%%%%%%%%%%%%%%%%%%%%%%%%%%%%%%%%%%%%%%

Over the past decade, there has been major progress in novel nonlinear photonic integrated platforms, including \SiN\cite{Xuan:16, Ji:17, Ye:19b, Liu:21}, AlGaAs\cite{Pu:16, Chang:20}, GaP\cite{Wilson:20}, tantala\cite{Belt:17, Jung:21}, and chalcogenide\cite{Eggleton:11, Liu:19, Kim:20}. 
These integrated platforms exhibit wide bandgaps and significantly higher effective nonlinearity than that of silica fibers, and allow lithographically tailored dispersion. 
Yet, a continuous-wave-pumped TWPA in the optical domain - i.e. a TWOPA - has remained out of reach using nonlinear photonic integrated circuits. 
Net gain has only been achieved using pulsed optical pump fields, to overcome the large optical losses of waveguides. 
%A continuous-wave TWOPA, capable of amplifying arbitrary temporal input signals, has to date been challenging using nonlinear photonic integrated circuits. 
A continuous-wave TWOPA, capable of amplifying arbitrary temporal input signals, has to date not been demonstrated with net gain using nonlinear photonic integrated circuits. 

Here we overcome this challenge and demonstrate a chip-based TWOPA that operates in the continuous-wave regime and achieves net-gain fiber-to-fiber. 
Our work is based on recent advances of ultralow-loss, dispersion-engineered, nonlinear, Si$_3$N$_4$ integrated waveguides that are fabricated using an optimized photonic Damascene process\cite{Pfeiffer:18b, Liu:21}.
Stoichiometric Si$_3$N$_4$ exhibits a transparency window from the visible to mid-infrared, and a bandgap of 5~eV that prohibits two-photon absorption in the 1550 nm band. 
It can be deposited via chemical vapor deposition and is CMOS-compatible\cite{Levy:10, Moss:13}.  
In contrast, the pioneering breakthrough of Foster \textit{et al}.\cite{Foster:06} has achieved  4.2~dB on/off gain in a silicon-on-insulator (SOI) waveguide, despite the two-photon absorption of silicon in the telecommunication bands, by using a picosecond pulsed laser for amplification to achieve high pump peak power.  
Most of state-of-the-art works\cite{Lamont:08, Kuyken:11, Wang:15_SiH, Ooi:17, liu2010mid, gajda2012highly} follow a similar scheme using pulsed pump lasers rather than continuous-wave lasers.

Continued and significant advances to reduce waveguide losses in integrated photonics\cite{Xuan:16, Ji:17, Ye:19b, Liu:21} over the past decade now culminate into a shift of this general paradigm.
As shown here, these advances allow efficient parametric generation and amplification in integrated nonlinear waveguides without the need for low-duty cycle pumping\cite{Foster:06,Lamont:08,Kuyken:11} or resonant enhancement\cite{Morichetti:11}. 
Time- and spectrum-continuous travelling-wave amplification is pivotal for successful implementation of amplifier technologies in modern optical communication systems, as well as emerging applications such as LiDAR.  
Numerous prior studies have reported on the progress of chip-based TWOPA and investigated new materials such as hydrogenated amorphous silicon (a-Si:H)\cite{Wang:15_SiH}, AlGaAs\cite{Pu:18}, and silicon-rich nitride (Si$_7$N$_3$)\cite{Ooi:17}. 
The performance of SOI waveguide systems has also been improved by operation in the mid-infrared region\cite{liu2010mid,Kuyken:11} or by active extraction of generated photocarriers in a PIN junction\cite{gajda2012highly}. 
The main focus of these works is the improvement of the so-called nonlinear figure-of-merit, i.e. the relation between the Kerr nonlinearity and the nonlinear absorption by careful balance of the electronic bandgaps and pump wavelengths.
Recent advances in fabrication\cite{Xuan:16, Ji:17, Ye:19b, Liu:21} have achieved crack-free \SiN photonic integrated circuits featuring tight optical confinement, high peak and average power handling capability, low Brillouin gain\cite{Gyger:20}, wideband engineering of anomalous group velocity dispersion~(GVD)\cite{Okawachi:11}, and ultralow optical losses near 1~dB/m with a $\chi^{(3)}$ nonlinear coefficient of up to 1~W$^{-1}$m$^{-1}$ and negligible nonlinear absorption at telecommunication bands. 
As such, optical spiral waveguides that confine pump and signal light over meter-long distances and boost the nonlinear interaction to levels unattainable before in \SiN waveguides become manufacturable now\cite{Liu:21}, and build the foundation of this work.

%%%%%%%%%%%%%%%%%%%%%%%%%%%%%%%%%%%%%%%%%%%%%%%%%%%%%%%%%%%%%%%%%%%%%%
%%%%%%%%%%%%%%%%%%% T H E O R Y   S E C T I O N %%%%%%%%%%%%%%%%%%%%%%
%%%%%%%%%%%%%%%%%%%%%%%%%%%%%%%%%%%%%%%%%%%%%%%%%%%%%%%%%%%%%%%%%%%%%%

Single-pump parametric amplification can be described using a frequency-domain model of waveguide modes coupled through nonlinear degenerate four-wave mixing (DFWM) mediated by the optical Kerr effect\cite{Stolen:82,Hansryd:02}, as the general principle is shown in Fig. \ref{Fig:1}(a,b). 
%Figure \ref{Fig:1} presents the general principle for a single-pump TWPA based on the optical Kerr effect \cite{Stolen:78,Hansryd:02}. 
A signal and a strong pump are combined and coupled into an optical waveguide, where power is transferred from the pump to the signal via DFWM. 
For every annihilated pair of pump photons of frequency $\omega_\text{P}/2\pi$, a signal photon $\omega_\text{S}/2\pi$ is generated together with a phase-conjugated idler photon $\omega_\text{I}/2\pi$, i.e. $2\omega_\text{P}=\omega_\text{S}+\omega_\text{I}$. 
In the absence of optical propagation loss, the signal power $P_\textrm{S}(L)$ and idler power $P_\textrm{I}(L)$ at the end of the waveguide of length $L$ follow as: 
\begin{equation}
\begin{gathered}
P_\textrm{S}(L) = P_\textrm{S}(0) \left( 1 + \left[ \dfrac{\gamma P_\textrm{P}(0)}{g} \sinh{gL}\right]^2\right)  \\
P_\textrm{I}(L) = P_\textrm{S}(0) \left[ \dfrac{\gamma P_\textrm{P}(0)}{g}\sinh{gL}\right]^2
\end{gathered}
\end{equation}
where $P_\textrm{S}(0)$ and $P_\textrm{P}(0)$ are the incident power of the signal and the pump. 
%what is $\gamma$
The parametric gain coefficient is derived as $g = -\Delta\beta \left( \Delta\beta/4  + \gamma P_\textrm{P}\right)$, with $\gamma$ being the coefficient describing the effective Kerr nonlinearity.  
The coherent nature of parametric interaction in the waveguide demands to fulfil a stringent phase matching condition for efficient amplification: 
\begin{equation}
\begin{gathered}
\Delta \beta = 2\beta(\omega_\textrm{P}) - \beta(\omega_\textrm{S}) - \beta(\omega_\textrm{I}) \\ 
\approx \beta_2\left(\omega_\textrm{S}-\omega_\textrm{P}\right)^2 + \dfrac{\beta_4}{12}\left(\omega_\textrm{S}-\omega_\textrm{P}\right)^4. 
\end{gathered}
\end{equation}
where $\beta$ denotes the optical propagation constant, and $\beta_2$ and $\beta_4$ are the second- and fourth-order derivative with respect to $\omega$. 
The GVD parameter $\beta_2$ of integrated waveguides can be engineered over a wide range by variation of waveguide cross-sectional geometry\cite{Turner:06,Okawachi:11}. 
As such, both fiber- and waveguide-based TWOPA systems can provide amplification bandwidths that greatly exceed those of rare-earth-doped fiber amplifiers. 
Figure \ref{Fig:1}(c) depicts the maximum achievable amplification bandwidth of a 2-meter-long \SiN waveguide spiral with a cross-section of $2.1\times0.67$~$\mu$m$^2$ (optimized for $\beta_2$ close to zero at the pump of 1550 nm) that can be fabricated on a $5\times5$~mm$^2$ size silicon chip. 
Such a system could exceed the EDFA's C-band amplification bandwidth (black bar in Fig. \ref{Fig:1}(c)) by more than three times. 
The amplification bandwidth can be reduced by increasing the waveguide height, leading to increased anomalous GVD. 

%%%%%%%%%%%%%%%%%%%%%%%%%%%%%%%%%%%%%%%%%%%%%%%%%%%%%%%%%%%%%%%%%%%%%%
%%%%%%%%%%%%%%%%%%% T E C H N I C A L   P A R T %%%%%%%%%%%%%%%%%%%%%%
%%%%%%%%%%%%%%%%%%%%%%%%%%%%%%%%%%%%%%%%%%%%%%%%%%%%%%%%%%%%%%%%%%%%%%

The \SiN photonic chip used in this work is shown in Fig. \ref{Fig:1}(d), and a microscope image of a corner of the waveguide spiral is shown in Fig. \ref{Fig:1}(e). 
Figure \ref{Fig:1}(f) depicts a scanning electron micrograph (SEM) of the chip cross-section, showing two \SiN waveguide cores of 0.91~$\mu$m height and 2.45~$\mu$m width. 
The mode profiles of the fundamental transverse-electric modes (TE$_{00}$) are superimposed on the waveguide cores. 
We measure the spiral waveguide's transmission spectrum, dispersion profile and propagation loss with a customized, polarization-maintaining, scanning, diode laser spectrometer and frequency-domain reflectometer (OFDR) in the wavelength range from 1260 to 1630~nm, calibrated using a self-referenced fiber-laser frequency comb\cite{Liu:16}.   
The results are presented in the Supplementary Material. 
The optical transmission through the 2-meter-long spiral is measured as high as 12$\%$ with a mean transmission of 10$\%$ in the wavelength range between 1545~nm and 1557~nm. 
The optical propagation losses are measured as low as 2.5~dB/m with a fiber-to-chip coupling loss of 2.5~dB per facet using lensed fibers and inverse tapers\cite{Liu:18}. 
The measured anomalous GVD is $-134$~fs$^2$mm$^{-1}$, in good agreement with the result from finite-element modelling based on the measured waveguide cross-section.

The experimental setup to measure the parametric gain and frequency conversion efficiency is depicted in Fig.~\ref{Fig:2}(a). 
Pump and signal lasers are derived from external-cavity diode lasers. 
We use a high-power EDFA and a tunable bandpass filter to amplify the pump and to remove excess amplified spontaneous emission (ASE) noise. 
The pump and signal are combined on a fused fiber beam splitter. 
The signal power coupled into the photonic chip is kept below 200~$\mu$W to avoid pump depletion. 
The signal gain and idler conversion efficiency are obtained from two signal laser wavelength scans performed in opposite direction starting at the pump wavelength and recorded on an optical spectrum analyzer (OSA) using the max-hold mode. 
Details are presented in the Supplementary Material. 
The transmission of the power-amplified pump laser is carefully optimized to 12\% in agreement with the calibrated transmission measurement at low optical power. 
The total fiber-to-fiber loss including the fiber-chip coupling losses (two facets) and optical propagation loss in the \SiN waveguide spiral is marked as grey dotted lines in Fig.~\ref{Fig:2}(b,d) and reaches as low as 10~dB. 
Therefore, we achieve for the first time a net parametric gain of up to 2~dB on a photonic chip accounting for both the on-chip optical propagation loss and the fiber-chip-fiber coupling losses. 
In addition, no damages of the waveguide and coupling facets are observed at input power levels up to 7~W. 
Notably this gain and power level are sustained without any mitigation techniques for stimulated Brillouin scattering such as fast pump laser dithering\cite{blows2002low} or phase modulation\cite{mussot2004impact}. 

%The results are depicted in Figure~\ref{Fig:2} (b,d). 
As depicted in Fig.~\ref{Fig:2}(b,d), the measured full bandwidths of gain and frequency conversion reach 20~nm despite the significant anomalous GVD of our thick \SiN waveguide. 
The measurement results are commensurate with our numerical calculations as shown in Fig.~\ref{Fig:2}(c,e), using the full set of nonlinear equations in the frequency domain\cite{Hansryd:02}. 
Notably, the literature value\cite{ikeda2008thermal} widely cited for \SiN nonlinear refractive index of $n_2=2.4\times 10^{-19}$~m$^2$W$^{-1}$ would result in a peak signal gain $G_\textrm{S}$ in excess of 18~dB. 
Recent measurements\cite{Gao:21} of the \SiN nonlinear refractive index reveal a reduced value of $n_2=2.1\times 10^{-19}$~m$^2$W$^{-1}$, likely due to a reduced fraction of Si-Si and Si-H bonds in high-temperature grown and annealed stoichiometric \SiN used for low absorption loss. 
We estimate the effective mode area $A_\textrm{eff}$ as small as $1.67$~$\mu$m$^2$ and the effective nonlinearity $\gamma$ of our waveguide as 0.51~W$^{-1}$m$^{-1}$, respectively. 
%While a reduction of the waveguide cross-section would allow to increase the effective nonlinearity up to 1~W$^{-1}$m$^{-1}$, it comes at a cost of increased scattering losses at the waveguide sidewall a requires substantial improvements in fabrication processes to reach the fundamental absorption loss limit of \SiN\cite{Liu:21}. 
With these parameters, our numerical calculations predict a peak gain of 12~dB in good agreement with measurements. 
Fluctuations of the waveguide cross-section that flatten and broaden the parametric gain lobes, and the remaining uncertainty around the transmission loss value\cite{karlsson1998four}, are negligible in the strong anomalous GVD regime of our waveguide.

%%%%%%%%%%%%%%%%%%%%%%%%%%%%%%%%%%%%%%%%%%%%%%%%%%%%%%%%%%%%%%%%%%%%%%
\begin{figure}[t!] 
\includegraphics[width=\linewidth]{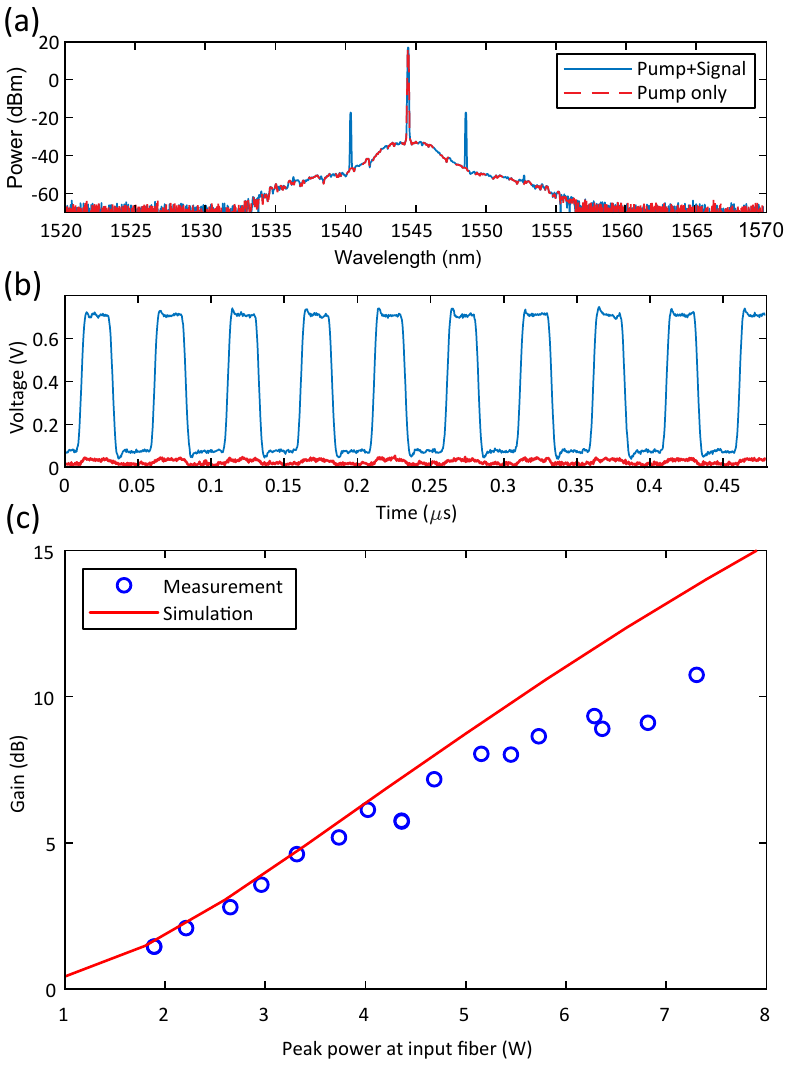}
\caption{
\textbf{Parametric gain measurement using pump modulation}.
(a)~Optical spectrum showing the pump, signal, and idler. 
(b)~Measurement of signal modulation due to parametric gain (blue). 
Small modulation observed without the signal (red) is obtained from modulation instability of the strong pump laser and finite bandwidth of the bandpass filter used to reject the pump light.
(c)~Measured optical signal gain extracted from the modulation measurement (blue). 
A simulated gain curve is depicted in red.
}
\label{Fig:3}
\end{figure}
%\vspace{-1mm}
%%%%%%%%%%%%%%%%%%%%%%%%%%%%%%%%%%%%%%%%%%%%%%%%%%%%%%%%%%%%%%%%%%%%%%

As an independent check, we also measure the parametric gain by fast modulation of the pump laser. 
We modulate the pump laser amplitude with a 50~MHz square wave before amplification with a duty cycle of 50\%. 
The instantaneous nature of parametric amplification mediated by the optical Kerr effect imprints the pump modulation directly on the amplified signal and generated idler. 
The measurement results are depicted in Fig.~\ref{Fig:3}. 
The pump laser is tuned to 1544.5~nm for this measurement, and the signal laser is tuned to 1548.5~nm. 
A second low-loss spiral is used for the measurement with similar waveguide design and from the same wafer substrate. 
The modulated signal beam is filtered using a tunable optical bandpass filter and is recorded on a fast photodetector. 
Figure \ref{Fig:3} depicts the extracted gain values as a function of the input peak power that is seen at the tip of the lensed fiber used to couple light into the chip. 
The results are in agreement with numerical calculations of the gain up to a peak input power of 5~W where we see a deviation up to 2.5~dB. 
Still, the parametric gain is able to compensate the total losses of our photonic chip. 

%%%%%%%%%%%%%%%%%%%%%%%%%%%%%%%%%%%%%%%%%%%%%%%%%%%%%%%%%%%%%%%%%%%%%%
%%%%%%%%%%%%%%%%%%%%%%% C O N C L U S I O N  %%%%%%%%%%%%%%%%%%%%%%%%%
%%%%%%%%%%%%%%%%%%%%%%%%%%%%%%%%%%%%%%%%%%%%%%%%%%%%%%%%%%%%%%%%%%%%%%

In summary, we present for the first time a continuous-travelling-wave optical parametric amplifier using a 2-meter-long \SiN nonlinear integrated waveguide that is capable to provide net gain off-chip. 
This advancement mirrors the breakthroughs in fiber-based rare-earth-doped\cite{Mears:87,Desurvire:87} and parametric\cite{Hansryd:01} amplifier technologies. 
The transition from fibers to nonlinear photonic integrated circuits bears great potential to improve overall device performance, footprint, and design freedom. 
Using advanced dispersion-engineering techniques such as vertically coupled waveguides\cite{zhang2010flattened} could significantly extend the parametric bandwidth by eliminating the second- and fourth-order dispersion terms in the phase mismatch.  
The fundamental absorption loss\cite{Liu:21} of stoichiometric \SiN waveguides was measured as low as 0.15~dB/m, which would facilitate parametric gain above 70~dB according to our simulations using as little as 500~mW of optical pump power, exceeding the performance of the best fiber-based parametric amplifiers\cite{torounidis2006fiber}. 
Moreover, the high nonlinearity and superior Kerr-to-Brillioun gain ratio\cite{Gyger:20} of \SiN waveguides will enable singly-resonant TWOPA with pump enhancement, which requires less footprint on chip, and greatly reduce the detrimental signal-signal FWM interaction that impeded the widespread adoption of fiber-based TWOPA originally developed in the late 1990s\cite{Marhic:15}. 
Such integrated-photonics-based TWOPAs would be directly pumped by high-power semiconductor lasers, inherently uni-directional, high-gain, broadband, as well as time- and spectrum-continuous with a quantum-limited noise figure,. 
They have the potential to become long-pursued candidates for future generations of optical communication systems that operate on the full transmission bandwidth of optical fibers\cite{Winzer:18}. 

%\textit{Note added}: During the preparation of this manuscript, we became aware of a talk presetend by researchers at Chalmer University of Technology during CLEO US 2021}. 

%\medskip

\noindent \textbf{Methods}

\begin{footnotesize}
\noindent \textbf{Fabrication process}: 
The \SiN photonic chips are fabricated using an optimized photonic Damascene process\cite{Pfeiffer:18b, Liu:21}. 
The waveguide pattern are written by deep-ultraviolet (DUV) stepper lithography based on 248~nm KrF excimer laser.
Advantages of using DUV stepper lithography instead of electron-beam lithography include smaller and fewer field stitching errors, easy implementation of multipass writing, and high fabrication throughput.
The patterns are dry-etched to the SiO$_2$ substrate to create waveguide preforms.
The substrate is then annealed at 1250$^\circ$C (``preform reflow'')\cite{Pfeiffer:18} to further reduce the root mean square (RMS) roughness of the waveguide sidewalls to sub-nanometer level.
Stoichiometric \SiN film of around 1$\mu$m thickness is deposited on the patterned substrate via low-pressure chemical vapor deposition (LPCVD), and fill the preform trenches to form the waveguide cores. 
An etchback planarization process\cite{Liu:21}, combining dry etching and chemical-mechanical planarization (CMP), is used to remove excess \SiN and to create waveguide top surface with 0.3~nm RMS roughness.
Afterwards, the substrate is annealed at 1200$^\circ$C with nitrogen atmosphere to drive out the residual hydrogen impurities in the \SiN film\cite{Liu:18a, Luke:15}.
A top SiO$_2$ cladding composed of TEOS and low-temperature oxide is deposited on the wafer, followed by SiO$_2$ annealing at 1200$^\circ$C.
Finally, the wafer is separated into individual photonic waveguide chips via deep dry etching followed by dicing or backside grinding. 

\noindent \textbf{Funding Information}: 
This work was supported by the Air Force Office of Scientific Research (AFOSR) under Award No. FA9550-19-1-0250, 
by Contract HR0011-20-2-0046 (NOVEL) from the Defense Advanced Research Projects Agency (DARPA), Microsystems Technology Office (MTO), 
by the Swiss National Science Foundation under grant agreement No. 176563 (BRIDGE), and by the EU H2020 research and innovation programme under grant agreement No. 965124 (FEMTOCHIP). 
J.R. acknowledges support from the EUs H2020 research and innovation program under the Marie Sklodowska-Curie IF grant agreement no. 846737 (CoSiLiS).

\noindent \textbf{Acknowledgments}: 
We thank Tianyi Liu for helping the design, and Miles H. Anderson for the discussion. 
The Si$_3$N$_4$ chips were fabricated in the EPFL center of MicroNanoTechnology (CMi). 

\noindent \textbf{Author contributions}: 
J.L. and R.N.W. designed and fabricated the samples. 
J.R., N.K. and J.H. performed the experiments and data analysis.
J.R. performed the numerical simulations. 
J.R., J.L. and T.J.K. wrote the manuscript. 
T.J.K. supervised the project.

%\noindent \textbf{Data Availability Statement}: 
%The code and data used to produce the plots within this work will be released on the repository \texttt{Zenodo} upon publication of this preprint.

\end{footnotesize}
%\vspace{-0.3cm}
\bibliographystyle{apsrev4-1}
\bibliography{bibliography}
\end{document}

\title{Photonic chip-based continuous-travelling-wave parametric amplifier}

\author{Johann Riemensberger}
\email[]{johann.riemensberger@epfl.ch}
\affiliation{Institute of Physics, Swiss Federal Institute of Technology Lausanne (EPFL), CH-1015 Lausanne, Switzerland}

\author{Junqiu Liu}
\affiliation{Institute of Physics, Swiss Federal Institute of Technology Lausanne (EPFL), CH-1015 Lausanne, Switzerland}

\author{Nikolai Kuznetsov}
\affiliation{Institute of Physics, Swiss Federal Institute of Technology Lausanne (EPFL),
CH-1015 Lausanne, Switzerland}
\affiliation{Russian Quantum Center (RQC), 143026 Skolkovo, Russia}
\affiliation{Moscow Institute of Physics and Technology (MIPT), Dolgoprudny, Moscow Region 141700, Russia}

\author{Jijun He}
\affiliation{Institute of Physics, Swiss Federal Institute of Technology Lausanne (EPFL), CH-1015 Lausanne, Switzerland}

\author{Rui Ning Wang}
\affiliation{Institute of Physics, Swiss Federal Institute of Technology Lausanne (EPFL), CH-1015 Lausanne, Switzerland}

\author{Tobias J. Kippenberg}
\email[]{tobias.kippenberg@epfl.ch}
\affiliation{Institute of Physics, Swiss Federal Institute of Technology Lausanne (EPFL), CH-1015 Lausanne, Switzerland}

\maketitle

%%%%%%%%%%%%%%%%%%%%%%%%%%%%%%%%%%%%%%%%%%%%%%%%%%%%%%%%%%%%%%%%%%%%%%
%%%%%%%%%%%%%%%%% C H A R A C T E R I Z A T I O N   %%%%%%%%%%%%%%%%%%
%%%%%%%%%%%%%%%%%%%%%%%%%%%%%%%%%%%%%%%%%%%%%%%%%%%%%%%%%%%%%%%%%%%%%%

\section{Characterization of waveguide spirals}

\noindent We measure transmission, reflection, dispersion and propagation loss of meter-long \SiN waveguide spirals with a custom frequency-comb calibrated scanning diode laser spectrometer. The optical setup and results for the waveguide spiral used in the experiment that was depicted in Figure 2 of the main manuscript is depicted in \ref{SI:Fig:1}. It is an extension of our earlier instrument \cite{Liu:16}, which was used successfully for the measurement of optical microresonators, their loss and dispersion. Three wideband mode-hop free tunable external-cavity diode lasers (ECDL,~Santec 710) cover the wavelength ranges between 1260-1360~nm (green), 1350-1505~nm (red), and 1500-1630~nm (blue) and are operated in sequence. Regular calibration markers are recorded by filtering a beat note between the lasers and a commercial optical freqeuncy-comb (Menlo~OFC-1500) using a balanced photoreceiver and logarithmic RF detector (LA, AD8307) for increased dynamic range. Furthermore, an imbalanced fiber-optical Mach-Zehnder Interferometer (MZI) and a molecular gas cell are used for further calibration and wavelength determination. 

%%%%%%%%%%%%%%%%%%%%%%%%%%%%%%%%%%%%%%%%%%%%%%%%%%%%%%%%%%%%%%%%%%%%%%
%%%%%%%%%%%%%%%%%%%%%%%%%% F I G U R E 1  %%%%%%%%%%%%%%%%%%%%%%%%%%%%
%%%%%%%%%%%%%%%%%%%%%%%%%%%%%%%%%%%%%%%%%%%%%%%%%%%%%%%%%%%%%%%%%%%%%%
\begin{figure*}[ht]
\centering
\includegraphics[width=\linewidth]{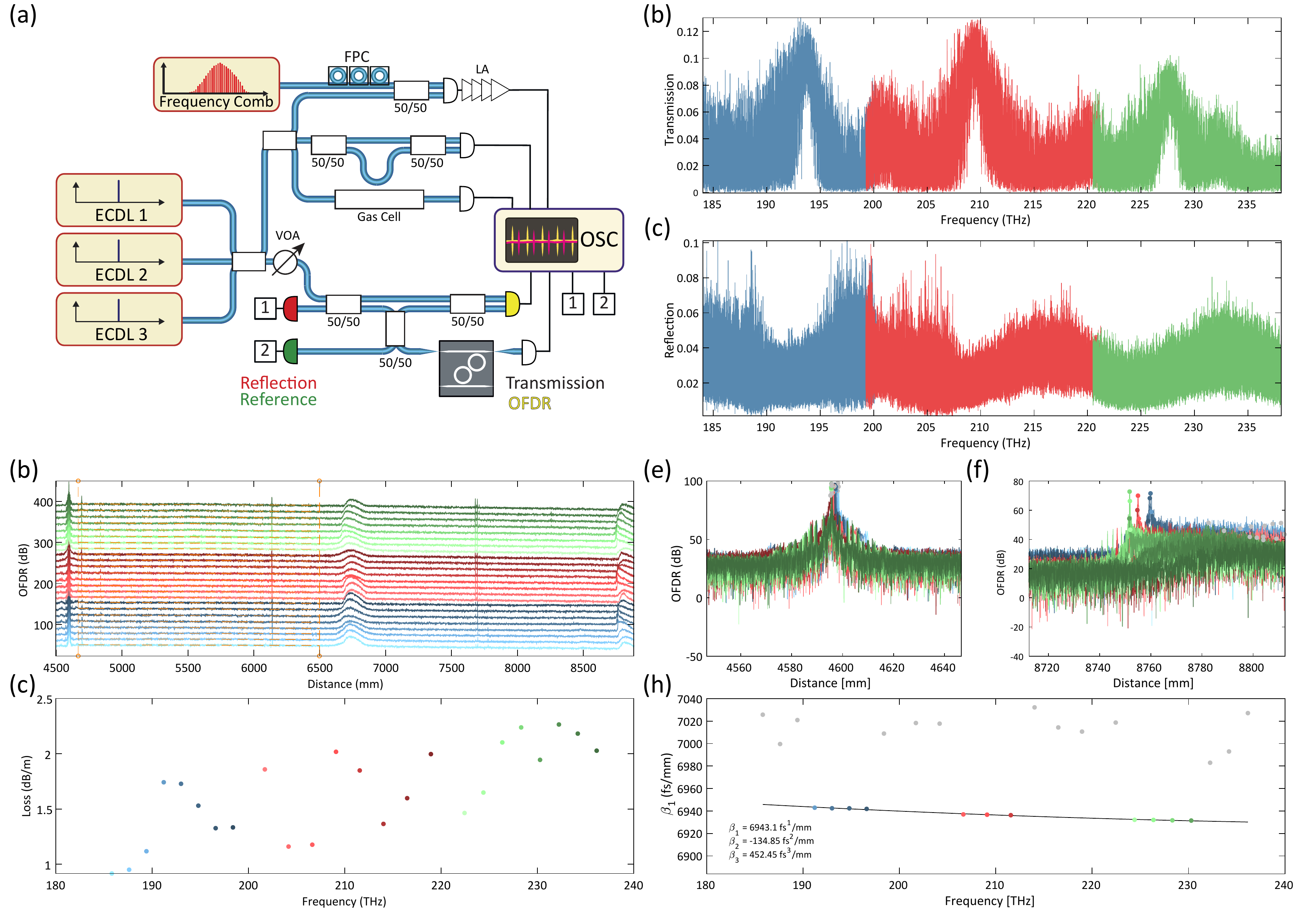}
\caption{\textbf{Frequency-comb calibrated characterization of waveguide spirals}.
(a)~Optical setup. See text for description.
(b)~Calibrated transmission spectrum. The trace colors indicate the laser ECDL laser number.
(c)~Calibrated reflection spectrum.
(d)~Optical frequency domain reflectrometry (OFDR). Traces are analyzed using segmented Fourier transform and vertically offset by 15 dB. The shading indicates the center frequency according to panel (e). The propagation loss fitting region is marked with vertical orange lines. The propagation loss is determined from the dotted line fit.
(e)~Propagation loss extracted from OFDR. Extracted propagation losses are relative to optical distance and must be multiplied by the group index 2.08 for conversion to geometrical waveguide length. The values represent upper bounds due to a background of laser phase noise that induces an increased gradient.
(f,g) Zoom into OFDR trace around front(back) facet. Colored dots depict successful identification of backside facet reflection for valid dispersion measurement. Grey markers indicate 
(h) Inverse group velocity $\beta_1$ as function of wavelength. Markers correspond to (f,g). The black line indicates the fitted dispersion curve up to third order. 
}
\label{SI:Fig:1}
\end{figure*}
%%%%%%%%%%%%%%%%%%%%%%%%%%%%%%%%%%%%%%%%%%%%%%%%%%%%%%%%%%%%%%%%%%%%%%

\noindent Transmission, reflection, a input power reference, and optical frequency-domain reflectrometry (OFDR) traces are recorded at the four ends of a three-way MZI formed by three polarization maintaining (PM) 50/50 fiber beamsplitters (OZ~Optics). Light is coupled into and out of the photonic chip using PM lensed fibers (OZ~Optics). The traces are digitized using a 8 channel analog digital converter (OSC). Calibration of the transmission is performed by recording a trace, where the lensed fibers are replaced with a PM patch cord. Calibration of the spiral reflection trace is performed by replacing the input lensed fiber with a fiber-coupled PM retroreflector. Calibration is performed relative to the concurrently recorded input reference channel. The calibrated transmission and reflection spectra of the spiral is depicted in Figure \ref{SI:Fig:1}, respectively. The colors in Figure \ref{SI:Fig:1} denote the three laser scans. We attribute the significant modulation of the measured transmission to interferences from chip facet reflections, defect reflections and intermodal interference. We observe that the spiral transmits only in 1.5~Thz windows around 193~THz, 210~THz and 227~THz, where the transmission reaches up to 12\% and 10\% on average. Hence we attribute a mean loss level around 10~dB to the waveguide spiral in the relevant transmission window at 193~THz as is depicted in Fig. 2 (b,d) of the main manuscript. We attribute the existence of the transmission windows to a destructive interference of higher order mode generation in the two straight-to-circular transitions of each identical 90$^\circ$ corner of our rectangular waveguide spirals. Outside the transmission window, the light is coupled into higher order modes in the spiral corners. An increase in reflection in those areas hints towards the increased coupling of higher order modes to the nearest neighbor arms of the spiral before dissipation. We estimate a coupling loss of $\approx 4.5$~dB, nearly  fiber-to-fiber through the photonic chip and conclude a linear propagation loss in the transmission windows of 2.75~dB/m.

\noindent The OFDR traces are analyzed with segmented Fourier transform (see Figure~\ref{SI:Fig:1}~d,e,f) using eight segments per trace (24~in total) with a window overlap equal to one segment. Our precise frequency comb calibration entails that the distance axis of the measurement denotes the optical distance without the requirement to scale for a certain group velocity index or fiber interferometer reference as is usual with state-of-the-art OFDR instruments \cite{soller2005high}. The shading of the colored OFDR traces indicates the center frequency of a particular segment with darker color indicating a higher frequencies. The propagation losses are estimated by linear fitting of the distance-dependent optical backreflection. Due to an underlying frequency noise background due to laser phase noise, we interpret the measured values as an upper bound. The extracted values for the optical propagation loss are depicted in Figure~\ref{SI:Fig:1}~(g). We observe a mean defect density of 2~m$^{-1}$. The group velocity dispersion $\beta_2$ is evaluated from the frequency-dependent optical length of the spiral by identifying the main reflection peaks of the chip input and output facets. The physical length of the spiral is 2.0~m. Because the waveguide tapers are only 300~$\mu$m long, their contribution to the measured dispersion may be ignored. Due to the higher order mode coupling, we can only observe a sharp and coherent backreflection peak in the three transmission windows to extract the group velocity dispersion $\beta_2$, which we estimate to -134~fs$^2$/mm.

%%%%%%%%%%%%%%%%%%%%%%%%%%%%%%%%%%%%%%%%%%%%%%%%%%%%%%%%%%%%%%%%%%%%%%
%%%%%%%%%%%%%%%%%% G A I N   M E A S U R E M E N T  %%%%%%%%%%%%%%%%%%
%%%%%%%%%%%%%%%%%%%%%%%%%%%%%%%%%%%%%%%%%%%%%%%%%%%%%%%%%%%%%%%%%%%%%%

\section{Measurement of parametric gain and frequency conversion spectra}

\noindent The optical parametric gain spectra are determined as follows: Both the pump and signal lasers are derived from wideband continously tunable external-cavity diode lasers (Toptica CTL). The pump laser is amplified using a high power EDFA (Keopsys CEFA-C-PB-HP). We filter the amplified spontaneous emission light (ASE) using a tunable thin-film bandpass filter and combine the lasers on a~99/1~fused fiber beam splitter. The high power output from the splitter carries most of the pump light, while the signal light is attenuated by -20~dB in the splitter and generally kept below -200~$\mu$W throughout all measurements presented here. We perform two separate wavelength scans with the signal laser while the pump laser is kept at 1555~nm. The first lasers scan (Scan~1) is performed from the pump wavelength towards longer wavelength and the second scan (Scan~2) is performed towards lower wavelengths. The signal gain $G_\textrm{S}$ and idler frequency conversion efficiency $G_{I}$ are related to the power of the signal at the output of the waveguide. We use the Max-Hold function of the optical spectrum analyzer to record the full gain spectra at each power level in a single slow laser scan ($\approx$30~s) at a resolution bandwidth of 2~nm. The signal transmission and fiber-to-chip coupling loss spectra are measured in an independent measurement to derive the on-chip gain (only propagation loss) and off-chip net gain (including fiber-to-chip coupling loss) condition. We measure a transmission of the signal laser of -28~dB in absence of gain and frequency conversion due to the pump.

%%%%%%%%%%%%%%%%%%%%%%%%%%%%%%%%%%%%%%%%%%%%%%%%%%%%%%%%%%%%%%%%%%%%%%
%%%%%%%%%%%%%%%%%%%%%%%%%% F I G U R E 2  %%%%%%%%%%%%%%%%%%%%%%%%%%%%
%%%%%%%%%%%%%%%%%%%%%%%%%%%%%%%%%%%%%%%%%%%%%%%%%%%%%%%%%%%%%%%%%%%%%%
\begin{figure*}[ht]
\centering
\includegraphics[width=0.85\linewidth]{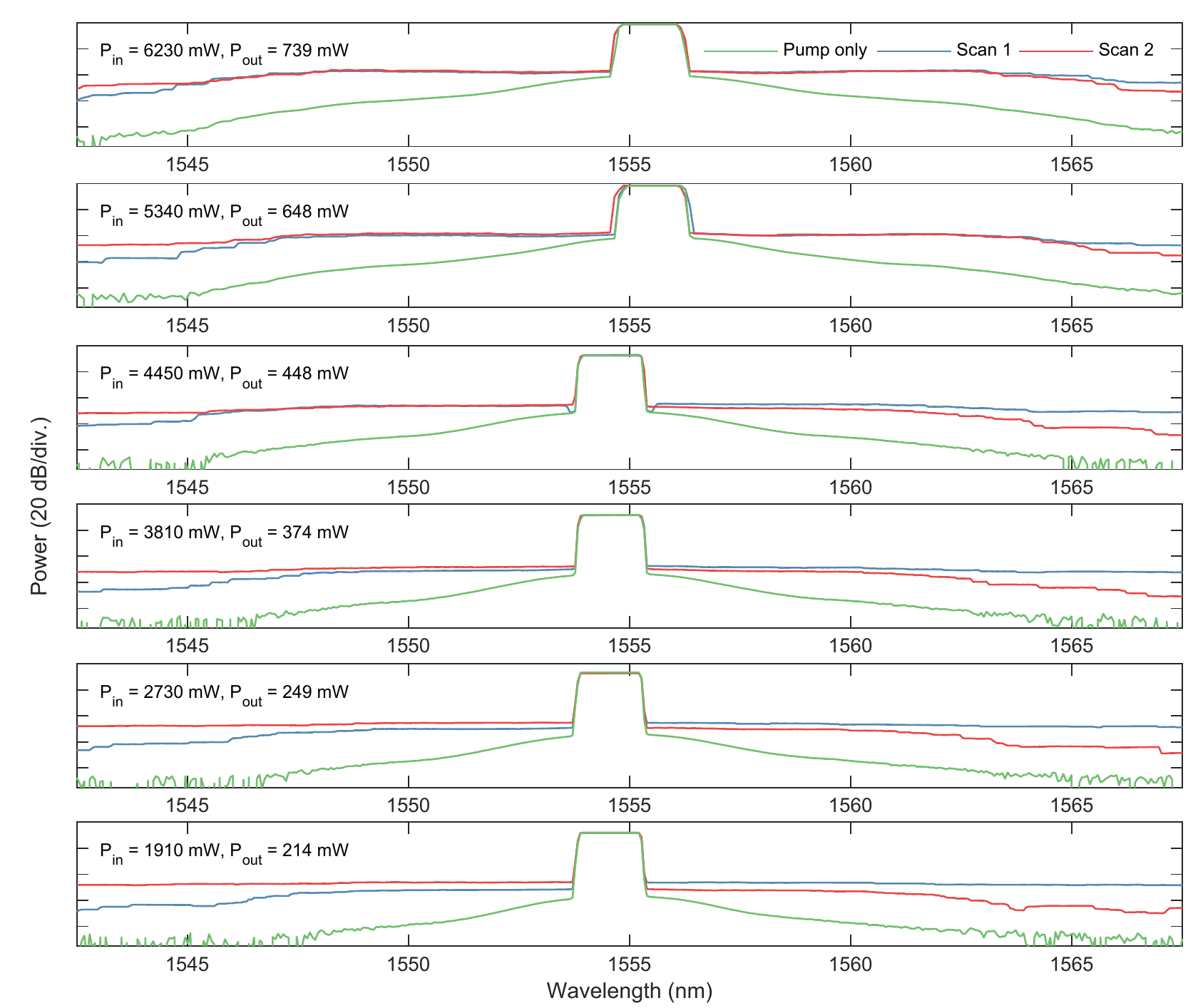}
\caption{\textbf{Raw data of gain spectrum measurement}. See text for detailed description.}
\label{SI:Fig:1}
\end{figure*}

%%%%%%%%%%%%%%%%%%%%%%%%%%%%%%%%%%%%%%%%%%%%%%%%%%%%%%%%%%%%%%%%%%%%%%
%%%%%%%%%%%%%%%% C O M P A R I S O N   T A B L E  %%%%%%%%%%%%%%%%%%%%
%%%%%%%%%%%%%%%%%%%%%%%%%%%%%%%%%%%%%%%%%%%%%%%%%%%%%%%%%%%%%%%%%%%%%%

\section{State of the art in planar waveguide TWOPA}

\noindent Signifcant progress has been made in the field of photonic integrated waveguide TWOPA's since the pioneering work of Foster et al. \cite{Foster:06}, who showed that a signal gain $G_\textrm{S}$ of up to 4~dB could be achieved in a silicon waveguide despite substantial two-photon absorption at telecom wavelength in the material. However, in order to achieve the high peak laser power required for this result, a picosecond pulsed pump laser was required to let the generated carriers dissipate before the next pulse and to not damage the waveguide with high average power. Since 2006, many groups have concentrated on the progress of integrated waveguide TWOPA investigating new materials such as hydrogenated amorphous silicon (a-Si:H) \cite{Wang:15_SiH}, aluminium gallium arsenide (AlGaAs)\cite{Pu:18}, ultra silicon rich nitride (Si$_7$N$_3$)\cite{Ooi:17}, or improving the performance of silicon-on-insulator waveguide systems by operation in the mid-infrared waveguide region \cite{liu2010mid,Kuyken:11} or by active extraction of generated photocarriers in a PIN junction \cite{gajda2012highly}. The main focus of this work is the improvement of the so-call nonlinear figure-of-merit, i.e. the relation between the Kerr nonlinear and the nonlinear absorption by careful balance of the electronic bandgaps and pump wavelengths. However, no continous-wave net gain on-chip or off-chip could be realized so far which is mostly due to the increased scattering and absorption losses of high refractive index materials and the added difficulty to couple broadband light into high-index contrast waveguides, often fabricated without top cladding. Simultaneously, high confinement silicon nitride waveguides were developed following the pioneering works of Gondarenko et al.\cite{Gondarenko:09} who overcame the high tensile stress of stochiometric \SiN waveguides and achieved the required thickness for anomalous waveguide dispersion. High waveguide scattering losses precluded the application of these early waveguides in TWOPA's. Driven by the desire to reduce the parametric oscillation threshold of soliton microcombs the waveguide losses were reduced by more than 1.5~orders of magnitude in the last 10 years, which now allows \SiN waveguides to achieve TWOPA operation with continuous-wave pump and net-gain on and off-chip for the first time. Table \ref{Tab1} summarizes the state of the art in the field. We estimated the total signal gain, idler conversion efficiency from the values reported in the literature, while summing up the reported losses from propagation and two fiber-to-chip coupling junctions.

%%%%%%%%%%%%%%%%%%%%%%%%%%%%%%%%%%%%%%%%%%%%%%%%%%%%%%%%%%%%%%%%%%%%%%
%%%%%%%%%%%%%%%%%%%%%%%% T A B L E 1 (to SI) %%%%%%%%%%%%%%%%%%%%%%%%%
%%%%%%%%%%%%%%%%%%%%%%%%%%%%%%%%%%%%%%%%%%%%%%%%%%%%%%%%%%%%%%%%%%%%%%

\begin{table}[ht]
\renewcommand{\arraystretch}{1.5}
\centering
\vspace{3mm}
\begin{tabularx}{\textwidth}{@{}
    >{\hsize=2\hsize\raggedright\arraybackslash}X 
    >{\centering\arraybackslash}X 
    >{\centering\arraybackslash}X
    >{\centering\arraybackslash}X
    >{\centering\arraybackslash}X
    >{\centering\arraybackslash}X
    >{\centering\arraybackslash}X
    >{\centering\arraybackslash}X
    >{\centering\arraybackslash}X
    >{\centering\arraybackslash}X
    >{\centering\arraybackslash}X
    >{\centering\arraybackslash}X@{}}
\toprule
Ref. & Material &  Year & $\lambda_\textrm{P}$ & $\gamma$    & $\alpha$ &   $\alpha_\textrm{tot}$ & G$_\textrm{S}$ & G$_\textrm{I}$ & P$_\textrm{pk}$ & D \\
~& ~& ~& (nm) & (1/Wm) &  (dB/m) & (dB)  & (dB) & (dB) &  (W) & (\%) \\
\midrule
% Foster 2006 Silicon
M.A. Foster et al.\cite{Foster:06}  & Si            & 2006  & 1550  & 155   & -120  & -26   & 4     & 5     & 11    & 0.03  \\ 
% Lamont As2S3
C. Lamont et al.\cite{Lamont:08}    & As$_3$S$_3$   & 2008  & 1532  & 9.9   & -50   & -17   & 32.5  & 30.6  & 9.6   & 0.003 \\ 
% Kuyken Silicon 
B. Kuyken et al.\cite{Kuyken:11}    & Si            & 2011  & 2173  & 150   & -280  & -26   & ~50   & ~50   &  13.5 & 0.02  \\ 
% Levy 2011
J. Levy\cite{Levy:11_PHD}           & \SiN          & 2011  & 1550  & 1.15  & -50   & -9    & 3.8   & 2     & 24    & 1     \\
% Gaija 2012 PIN junction silicon
A. Gadja et al.\cite{gajda2012highly} & Si          & 2012  & 1542  & 200   & -200  & -18   & 5    & ~3    & 0.4   & 100   \\
% Wang a-Si:H
K.Y.Wang et al.\cite{Wang:15_SiH}   & a-Si:H        & 2015  & 1550  & 0.64  & -35   & -19   & 12    & 10.5  & 0.75  & 0.16  \\
% OOi Si3N7 
K. Ooi et al.\cite{Ooi:17}          & Si$_7$N$_3$   & 2017  & 1550  & 500   & -45   & -17   & 42.5  & 36.2  & 14    & 0.001 \\
% Pu iii-V
M. Pu et al. \cite{Pu:18}           & AlGaAs        & 2018  & 1550  & 630   & -200  & -6.4  & 1.4  & -4.2  & 0.4   & CW    \\  
% This work
This work                           & \SiN          & 2021  & 1555  & 0.50  & -3    & -1.5  & 10    & 9     & 4.7   & CW    \\
\bottomrule
\end{tabularx}
\caption{\textbf{Comparison of the state of the art in planar integrated waveguide parametric amplifiers and frequency converters.} $\lambda_\textrm{P}$: pump wavelength; $\gamma$: effective nonlinearity; $\alpha$: linear propagation loss; $\alpha_\textrm{tot}$: total loss including coupling from and to optical fiber;  G$_\textrm{S}$: signal on-off gain; G$_\textrm{I}$: Frequency-conversion efficiency relative to signal output; P$_\textrm{pk}$: peak pump power; D: pump pulse duty cycle}
\label{Tab1}
\end{table}

\section*{Supplementary References}
\bigskip
\bibliographystyle{naturemag}
\bibliography{bibliography}